\documentstyle[preprint,aps,psfig]{revtex}
\newcommand{\beq}{\begin{equation}}
\newcommand{\eeq}{\end{equation}} 
\newcommand{\beqa}{\begin{eqnarray}}
\newcommand{\eeqa}{\end{eqnarray}} 
\newcommand{\ba}{\begin{array}} 
\newcommand{\ea}{\end{array}} 

\begin{document}
\draft


\widetext 
\title{Periodic Quantum Tunneling and Parametric Resonance\\ 
with Cigar-Shaped Bose-Einstein Condensates} 
\author{L. Salasnich$^{1}$, A. Parola$^{2}$ and L. Reatto$^{1}$} 
\address{$^{1}$Istituto Nazionale per la Fisica della Materia, 
Unit\`a di Milano, \\
Dipartimento di Fisica, Universit\`a di Milano, \\
Via Celoria 16, 20133 Milano, Italy\\
$^{2}$Istituto Nazionale per la Fisica della Materia, 
Unit\`a di Como, \\
Dipartimento di Scienze Fisiche, Universit\`a dell'Insubria, \\
Via Valeggio 11, 23100 Como, Italy}
\maketitle
\begin{abstract} 
We study the tunneling properties of a 
cigar-shaped Bose-Einstein condensate by using 
an effective 1D nonpolynomial nonlinear Schr\"odinger 
equation (NPSE). First we investigate a mechanism to generate 
periodic pulses of coherent matter by means 
of a Bose condensate confined in a potential well 
with an oscillating height of the energy barrier. We show that is 
possible to control the periodic emission of matter waves 
and the tunneling fraction of the Bose condensate. 
We find that the number of emitted particles strongly increases 
if the period of oscillation of the height of the energy barrier 
is in parametric resonance with the period of oscillation of 
the center of mass of the condensate inside the potential well. 
Then we use NPSE to analyze the periodic tunneling of a Bose-Einstein 
condensate in a double-well potential which has an oscillating 
energy barrier. We show that the dynamics of the Bose condensate critically 
depends on the frequency of the oscillating energy barrier. 
The macroscopic quantum self-trapping (MQST) of the condensate 
can be suppressed under the condition of parametric resonance between 
the frequency of the energy barrier and the frequency of oscillation 
through the barrier of the very small fraction of particles 
which remain untrapped during MQST. 
\end{abstract}

\vskip 1. truecm



\narrowtext

\newpage 

\section{Introduction}

In the last few years macroscopic quantum tunneling (MQT) 
with dilute Bose-Einstein condensates of alkali-metal atoms 
has been the subject of many theoretical [1-4] and experimental 
[5-7] studies. In a recent paper we have suggested a mechanism 
to generate pulses of coherent matter by using MQT 
of a falling and bouncing condensate [8]. 
In the first part of this paper we propose another, simpler, mechanism 
to produce periodic atomic wave-packets by means of MQT 
of a Bose condensate through the barrier of a potential well. 
In our theoretical investigation we consider cigar-shaped 
Bose condensates so that we can use an effective 1D nonpolynomial 
nonlinear Schr\"odinger equation, which we have shown to be quite accurate 
when the aspect ratio is larger than 3 [9,10]. 
We consider a Bose-Einstein condensate confined in the vertical 
axial direction by two Gaussian optical barriers and in the transverse 
direction by a magnetic or optical harmonic potential. By periodically 
changing the height of the lower-lying Gaussian barrier 
it is possible to generate periodic waves of 
coherent matter: only when the height of 
the energy barrier is sufficiently small a consistent 
fraction of condensed atoms can tunnel. We show that it is 
possible to control the period of emission of the matter waves 
and the tunneling probability. We find that 
the emission probability can be strongly enhanced by using 
the parametric resonance between the oscillation of the height 
of the energy barrier and the oscillation of the center of mass 
of the condensate inside the potential well 
(resonance-induced coherent emission). 
\par 
In the second part of the paper we study another problem of tunnelling, 
the one of trapped particles in a double potential well. In particular, 
we investigate the effect of a periodically varying 
barrier in the tunneling of the Bose condensate. 
The periodic oscillations of a Bose-Einstein condensate 
in a double-well has been the subject of many 
papers [1-4] but, to our knowledge, the role of a 
time-dependent energy barrier has not been analyzed. 
We investigate the problem using our effective 1D nonpolynomial 
Schr\"odinger equation and also the two-mode classical-like 
equations introduced by Smerzi et al [1]. The numerical results 
obtained with the two methods are in quantitative agreement. 
In particular, we find that the phenomenon of 
macroscopic quantum self-trapping of the condensate can be controlled 
(reduced or suppressed) by the frequency of oscillation of 
the double-well energy barrier again via parametric resonance. 

\section{Mapping from 3D to 1D} 

It is well know that the 3D Gross-Pitaevskii equation (GPE) 
is accurate in describing the Bose-Einstein condensate 
of a dilute boson gas at temperatures well below the 
transition temperature, where thermal 
excitations can be neglected [11]. 
Often the condensates produced in experiments are cigar-shaped and 
this suggests to map the 3D GPE into an effective 1D equation, 
which simplifies greatly the solution of the equation.  
This problem is not trivial due to the nonlinearity 
of the GPE. By using a variational approach, 
we have recently derived an effective time-dependent 
1D nonpolynomial nonlinear Schr\"odinger equation (NPSE) 
which describes very accurately cigar-shaped Bose condensates [9,10]. 
We assume that the condensate is confined by a magnetic (or optical) 
harmonic potential with frequency $\omega_{\bot}$ and harmonic 
length $a_{\bot}=(\hbar/m\omega_{\bot})^{1/2}$ 
in the transverse direction and by a generic potential $V(z)$ 
in the axial one. The total wave function of the condensate is 
$\psi(x,y,z,t) = f(z,t) g(x,y,t)$, where the transverse 
wave-function $g(x,y,t)$ is a Gaussian with the width $\eta$ 
such that $\eta^2=a_{\bot}^2\sqrt{1+2a_sN|f|^2}$ with $a_s$ 
the s-wave scattering length [10]. The NPSE is given by 
$$
i\hbar {\partial \over \partial t}f= 
\left[ -{\hbar^2\over 2m} {\partial^2\over \partial z^2} 
+ V(z) + {g N \over 2\pi a_{\bot}^2} 
{|f|^2\over \sqrt{1+ 2a_s N|f|^2} } 
\right. 
$$
\beq 
\left. 
+ {\hbar \omega_{\bot}\over 2}  
\left( {1\over \sqrt{1+ 2 a_s N|f|^2} } + \sqrt{1+ 2a_s N|f|^2}
\right) \right] f  \; , 
\eeq 
where $f(z,t)$ is the axial macroscopic wave function of the condensate, 
$g={4\pi \hbar^2 a_s/m}$ is the scattering amplitude and $N$ is the 
number of condensed bosons and the function $f(z,t)$ is normalized to one.  
\par
We observe that in the weakly-interacting limit $a_s N|f|^2 <<1$, 
NPSE reduces to a 1D GPE with the nonlinear coefficient $g'$ given 
by $g'=g/(2\pi a_{\bot}^2)$. Instead, in the strongly-interacting 
limit, NPSE becomes a nonlinear Sch\"odinger equation with 
the nonlinear term proportional to $|f|f$. 
\par 
In paper [10] we have tested the accuracy of the NPSE 
in the determination of the ground-state and collective 
oscillations of the condensate with axial harmonic 
confinement and also in the description 
of tunneling through a Gaussian barrier. 
In particular, we have compared NPSE with the the full 3D GPE. 
The conclusion is that NPSE is very accurate when the aspect ratio 
in larger than $3$ and gives better results than all other effective 
approaches recently proposed [12,13]. 

\section{Periodic emission of matter waves} 

In a previous paper [8] we have analyzed the 
pulsed quantum tunneling of a Bose condensate 
falling under gravity and scattering on a Gaussian 
barrier that model a mirror of far-detuned 
sheet of light. In this section we study another 
mechanism which produces periodic emission of 
matter waves by means of the quantum tunneling 
of a Bose condensate. We consider a cigar-shaped 
condensate confined by a harmonic potential in the 
transverse direction and under the action of 
the following external potential in the vertical 
axial direction: 
\beq 
V(z)= V_1 \; e^{-(z-z_1)^2/ \sigma^2} + 
V_2 \; e^{-(z-z_2)^2/ \sigma^2} + m g z \; . 
\eeq 
Thus, in addition to the gravity potential $mgz$, there are 
two Gaussian functions that model a confining potential well. 
The shape of this potential is shown in Figure 1. 
Such a configuration can be experimentally obtained by using  
two blue-detuned laser beams (perpendicular to 
the axial direction) which are modelled by the two Gaussian 
potentials. By varying the intensity of the lower-lying 
laser beam one controls the height of the energy barrier 
and therefore the tunneling probability. 
In particular we consider the case in which the height $V_2$ 
of the lower barrier is a sinusoidal function of time.  
As an example we consider a Bose-Einstein condensate of $^{23}$Na atoms 
with a scattering length $a_s=30$ $\AA$. 
For the frequency $\omega_{\bot}$ of the axial harmonic confinement 
we choose $\omega_{\bot} =2\pi$ kHz. Setting $\omega_z=\omega_{\bot}/10$, 
we measure lengths in units $a_z=(\hbar/m\omega_z)^{1/2}$, 
time in units $\omega_z^{-1}$ and energy in units $\hbar \omega_z$.   
In these units the period of small oscillations around the minimum 
of the potential well (2) close to one. 
The NPSE is integrated by using a finite-difference 
predictor-corrector method. 
The initial wave function of the condensate is found 
by solving the equation with imaginary time.  
\par 
First we calculate the ``ground-state'' of the system 
by choosing $V_1=1000$, $V_2=900$, $z_1=17$, 
$z_2=24$ and $\sigma=1.6$. Then we reduce the energy 
barrier $V_1$ studying the time evolution and tunneling 
of the condensate through the potential well. 
The results are shown in Figure 2. There is a periodic tunneling 
of matter waves and the period of tunneling is proportional 
to the period of small oscillations of the center of mass 
of the condensate around the minimum of the axial external 
potential $V(z)$. To compare the chemical potential of the 
Bose condensate with the energy barrier $V_1$, 
the effective chemical potential 
$\mu_{eff}$ of the Bose condensate is obtained from the full one 
by subtracting not only the transverse energy per particle 
but also the gravitational energy per particle. 
\par 
A simpler mechanism to generate periodic pulses is to 
have the lower-lying Gaussian barrier with a periodic height 
given by: 
\beq 
V_1(t)=V_a+V_b\sin{({2\pi\over \tau} t)} \; , 
\eeq 
where $\tau$ is the period of oscillation. 
We choose $V_a=300$ and $V_b=200$. 
In Figure 3 we plot the axial density profile $\rho(z)$ 
of the matter-waves coming out from the potential well 
with $\tau=1.5$. Figure 3 shows that, as expected, 
the emission of pulses is periodic. 
\par 
We find that both the period of emission and the tunneling 
fraction $P_T$ depend on $\tau$. 
For very small values of $\tau$, the fraction 
of ejected particles is quite small 
because the height of the energy barrier oscillates very quickly. 
On the other hand, for large values 
of $\tau$ the fraction $P_T$ is again very small. 
In the inset of Figure 4 the tunneling probability $P_T$ is shown 
as a function of $\tau$ for a Bose condensate 
with $N=10^3$ and $N=10^4$ atoms. 
The emission probability has its absolute maximum 
near $\tau=0.8$, that is the average period 
of the oscillations of the center of mass 
of the condensate inside the potential well. 
\par 
The emission probability $P_T$ grows by 
increasing the number $N$ of atoms and it is strongly 
enhanced by setting the oscillating barrier at the resonance 
condition. Nevertheless, by plotting the ratio $P_T/P_T^0$ 
as a function of $\tau$, where 
$P_T^0$ is the tunneling probability with $V_1(t)=V_a-V_b=100$, 
one finds that the resonance is suppressed by the interaction, 
i.e. by increasing $N$ (see Figure 4). 
Note that in the non-interacting case 
($N=0$ and $\mu_{eff}=21.88$) one finds 
$P_T^0=1.11\times 10^{-4}$, but $P_T=0.35$, thus 
the tunneling ratio $P_T/P_T^0$ is larger than 
$3\times 10^3$. 
\par 
The enhancement of the emission probability of coherent 
matter near $\tau=0.8$ has a classical explanation. 
It corresponds to the condition of resonance 
between the oscillation of the energy barrier and 
the oscillation of the center of mass of the condensate. 
Such an effect is clearly shown in Figure 5, where we plot the 
axial coordinate and the phase-space portrait of a classical particle 
under the action of the potential (2). 
The dynamics of the classical particle is obtained by solving the Hamilton 
equations with a 4th-order Runge-Kutta method. 
In Figure 5, for $\tau=0.75$ the system is close to the resonance 
condition: 
the periodic energy of the barrier is pumped into the energy of the 
classical particle and the particle is able to escape from the trap. 
Far from the resonance condition, the periodic energy of the barrier is not 
fully transferred to the energy of the classical particle, so the 
particle remains confined. 
Close to the resonance condition the phenomenon of coherent emission 
of atoms is no more due to quantum tunneling but to the parametric 
resonance [14] between the period of 
oscillation of the center of mass of the condensate 
and the period of oscillation of the energy barrier. 
\par
Because of the presence of an oscillating barrier of period $\tau$, 
the period ${\tilde\tau}_{osc}$ of oscillation of 
the center of mass of the condensate 
follows the law ${\tilde \tau}_{osc} = \tau_{osc} 
(1+\epsilon \sin(2\pi t/\tau))$, 
where $\tau_{osc}$ is the period of oscillation of the center 
of mass in the absence of an oscillating 
barrier and $\epsilon$ represents the amplitude of the periodic perturbation. 
Text-books [14] predict parametric resonance for 
$\tau = \left(n/2\right) \tau_{osc}$ with $n$ an integer number. 
In fact, as shown in Figure 4, 
there are other local maxima in the plot of the emission fraction 
whose positions follow this resonant condition formula. 

\section{Periodic tunneling in a oscillating double-well potential} 

In this section we investigate the dynamics of a Bose-Einstein 
condensate of $^{23}$Na atoms in a double-well trap given by a 
harmonic anisotropic potential plus a Gaussian 
barrier along the $z$ axis, which could model the effect of a laser 
beam perpendicular to the long axis of the condensate. 
In our model the height of the Gaussian barrier is periodic 
and performs small oscillations around its mean value. 
\par 
The Bose condensate is confined by a harmonic potential in the 
transverse direction while the external potential in the horizontal axial 
direction is given by: 
\beq 
V(z,t)= V_1(t) \; e^{-z^2/ \sigma^2} + {1\over 2} m \omega_z^2 z^2 \; , 
\eeq 
where again 
\beq 
V_1(t)=V_a+V_b\sin{({2\pi \over \tau}t)}  \; . 
\eeq 
We perform the numerical integration of NPSE with the potential (4) 
using the $z$-harmonic oscillator units with 
$\omega_z =\omega_{\bot}/10$ and $\omega_{\bot}=2\pi$ kHz. 
The shape of this double-well 
external potential is shown in Figure 6. 
Such a configuration can be experimentally obtained by using 
a procedure similar to that described in the previous section 
(see also [15]). 
\par
First we calculate the ``ground-state'' of the system fully confined 
in the right well by choosing $V_1=20$ and $V_2=0$. 
Then we reduce the energy barrier $V_1$ studying the time evolution 
and periodic tunneling of the condensate through the Gaussian barrier. 
In Figure 7 we plot the fraction $P_L$ of $^{23}$Na atoms 
in the left well as a function of time $t$, choosing $V_1=6.5$ and $V_2=0$. 
The behavior of $P_L(t)$ depends on the inter-atomic energy of the 
condensate. As shown in Figure 7, for small values of the inter-atomic 
strength $Na_s/a_z$ the condensed cloud oscillates between the two wells, 
while for larger values of $Na_s/a_z$ the condensate 
remains self-trapped. Thus, for large values of the inter-atomic 
interaction, only a small part of atoms oscillates 
between the two wells while the main part of the atomic condensate 
remains confined in the right well. 
This phenomenon, called macroscopic quantum self-trapping (MQST),  
has been recently predicted by Smerzi et al. [1]. 
\par 
Smerzi et al. [1] have found that the time-dependent behavior 
of the condensate in the tunneling 
energy range can be described by the two-mode equations 
\beq
{\dot \zeta}=-\sqrt{1-\zeta^2}\sin{\phi} \; , \;\;\;\;\;\;\; 
{\dot \phi}=\Lambda \zeta +{\zeta\over \sqrt{1-\zeta^2}}\cos{\phi} \; ,
\eeq 
where $\zeta=(N_1-N_2)/N$ is the fractional population 
imbalance of the condensate in the two wells, 
$\phi=\phi_1-\phi_2$ is the relative phase 
(which can be initially zero), and $\Lambda = 2 E_I/E_T$ with 
$E_I$ the inter-atomic energy and 
$E_T$ the tunneling energy, i.e. the kinetic+potential energy 
splitting between the ground state and the quasi-degenerate 
odd first excited state of the GP equation. Note that $E_T=\hbar \omega_0$, 
where $\omega_0$ is the oscillation frequency of the Bose condensate 
between the two wells when the inter-atomic interaction is zero ($E_I=0$). 
For a fixed $\Lambda$ ($\Lambda >2$), 
there exists a critical $\zeta_c =2\sqrt{\Lambda -1}/\Lambda$ 
such that for $0 < \zeta << \zeta_c$ there are Josephson-like 
oscillations of the condensate with period 
$\tau_J = \tau_0/\sqrt{1+\Lambda}$, where $\tau_0 =2\pi /\omega_0$. 
But for $\zeta_c<\zeta \leq 1$ there is macroscopic quantum 
self-trapping (MQST) of the condensate: even if the populations
of the two wells are initially set in an asymmetric state 
($\zeta(0)\ne 0$) they maintain, on the average, 
the original population imbalance with a very small periodic 
transfer of particles through the barrier with period 
$\tau_{ST}=\tau_0/\sqrt{2(\Lambda H_0-1)}$, where 
$H_0=\Lambda\zeta(0)^2/2-\sqrt{1-\zeta(0)^2}\cos{\phi(0)}$ [1]. 
Our NPSE numerical results of Figure 7 confirm the 
two-mode approximation (Eq. (6)) prediction: 
the fully imbalanced ($\zeta(0)=1$) Bose-Einstein condensate approaches 
a self-trapped regime by increasing $Na_a/a_z$, i.e. the inter-atomic energy. 
\par 
It is interesting to see what happens if we include a periodic oscillation 
of the barrier ($V_2\neq 0$). As shown in Figure 8, where we plot the 
condensate fraction $P_L(t)$ in the left-handed well obtained by numerically 
integrating the NPSE, the presence of an 
oscillating barrier of period $\tau$ can break or simply reduce the MQST of 
period $\tau_{ST}$. In particular, in Figure 8 it is shown that 
if $\tau=\tau_{ST}/2$ (parametric resonance condition [14]) 
then the MQST is broken; instead, if $\tau=2 \tau_{ST}$ 
the MQST is not broken 
but there is a larger periodic transfer of particles through the barrier. 
\par 
This phenomenon can be investigated by using the two-mode approximation 
of Eq. (4). In this case, the presence of a oscillating 
barrier can be modelled by a time-dependent $\Lambda$: e.g. 
$\Lambda(t) = \Lambda_0 (1+\epsilon \sin{(2\pi t/\tau)})$. 
In Figure 9 we plot the population imbalance 
$\zeta (t)$ and its phase-space portrait $(\zeta(t),{\dot \zeta}(t))$ 
with $\Lambda_0=25$, $\zeta (0)=0.6$, $\phi(0)=0$. Also in this case 
the MQST is strongly affected by the parametric resonance between 
the period $\tau$ of MQST oscillation and the period $T$ of oscillation 
of $\Lambda$ (i.e. the energy barrier). As shown in Figure 10, 
at the parametric resonance condition $\tau=\tau_{ST}/2$ 
and with a sufficiently large perturbation ($\epsilon =0.2$) 
the system escapes from the self-trapping 
configuration. We have verified that for this system the general 
parametric resonance condition $\tau=\left(n/2\right)\tau_{ST}$ [14] 
works quite well at small $n$ (even and odd).  
\par 
Recently the parametric excitation of cold trapped atoms 
in far-off-resonance optical lattices 
has been experimentally obtained by modulating the potential depth [16,17]. 
In these experiments the parametric resonance results in heating 
or losses for the trapped atoms and it can be studied to 
determine the spring constant of the periodic optical potential [17]. 
Another kind of parametric resonance, namely
the parametric resonance between the collective oscillations 
of a trapped Bose-Einstein condensate 
and the oscillations of the confining harmonic trapping potential 
has been demonstrated with numerical simulations 
by Kevrekidis, Bishop and Rasmussen [18]. They have shown that a 
weak harmonic modulation of the confining potential can cause an 
anomalously large amplitude in the collective oscillations of the Bose 
condensate. Actually, a similar effect can be obtained 
without parametric resonance by imposing particular values of anisotropy 
to the trapping potential: for such values different collective 
modes of the condensate are in resonance [19]. 

\section*{Conclusions} 

We have studied the macroscopic quantum tunneling of 
a cigar-shaped Bose-Einstein condensate confined 
by a harmonic potential in the transverse direction 
and tunneling through an energy barrier in the axial 
direction. The time-evolution of the macroscopic wave function 
of the condensate is obtained by using an effective 
1D nonpolynomial nonlinear Schr\"odinger equation. 
\par 
First we have investigated a Bose 
condensate confined in the vertical axial direction 
by two Gaussian barriers that model a potential well. 
By periodically changing the height of 
the lower-lying Gaussian barrier 
it is possible to generate periodic waves of 
coherent matter: only when the height of 
the energy barrier is sufficiently small a consistent 
fraction of condensed atoms can tunnel. We show that it is 
possible to control the period of emission of the matter waves 
and the tunneling probability. We have also found a 
resonance-induced coherent emission of atoms: 
the emission probability is strongly enhanced 
if the period of oscillation of the height 
of the energy barrier is in parametric resonance with period 
of oscillation of the center of mass 
of the condensate inside the potential well. 
\par 
Then we have studied the periodic tunneling and the quantum self-trapping 
of a Bose-Einstein condensate in a double-well 
potential with an oscillating energy barrier. 
We have used our effective 1D nonpolynomial Schr\"odinger equation 
and also the two-mode classical-like equations. 
We have found that the macroscopic quantum self-trapping 
of a Bose condensate can be reduced and also suppressed 
by changing the frequency of the oscillating barrier. 
In particular, the system escapes from the self-trapping configuration 
if the the period of oscillation of the double-well energy barrier and 
the period of MQST oscillations of the condensate 
satisfies the parametric resonance condition. 
\par 
The parametric driving of Bose-Einstein condensates can be 
obtained by current experiments using optical dipole forces with 
far-detuned laser beams [20]. 
By varying the intensity of the laser beams one can control 
the height of the energy barrier that confines the condensed sample. 
The experimental investigation of the 
physical configurations we have considered in this paper 
may contribute to the realization of novel phenomena: 
controlled periodic emission in atom lasers and 
breaking of macroscopic quantum self-trapping 
induced by parametric resonance. 

\newpage

\section*{References}

\begin{description}

\item{\ [1]} Smerzi A, Fantoni S, Giovannazzi S, 
and Shenoy S R 1997 Phys. Rev. Lett. {\bf 79} 4950;  
Raghavan S, Smerzi A, Fantoni S, 
and Shenoy S R 1999 Phys. Rev. A {\bf 59} 620. 

\item{\ [2]} Milburn G J, Corney J, Wright E, and 
Walls D F 1997 Phys. Rev. A {\bf 55} 4318  

\item{\ [3]} Zapata I, Sols F, and Leggett A J 1998 
Phys. Rev. A {\bf 57} R28 

\item{\ [4]} Salasnich L, Parola A, and Reatto L 1999  
Phys. Rev. A {\bf 60} 4171; 
Pozzi B, Salasnich L, Parola A, and Reatto L 2000  
Eur. Phys. J. D {\bf 11} 367 

\item{\ [5]} B.P. Anderson and M. Kasevich 1999 Science {\bf 282} 
1686 

\item{\ [6]} Cataliotti F S, Burger S, Fort C, Maddaloni P, 
Trombettoni A, Smerzi A, and Inguscio M 
2001 Science {\bf 293} 843 

\item{\ [7]} Morsch O, Muller J H, Cristiani M, Ciampini D, and 
Arimondo E 2001 Phys. Rev. Lett. {\bf 87} 140402 

\item{\ [8]} Salasnich L, Parola A, and Reatto L 2001  
Phys. Rev. A {\bf 64} 023601 

\item{\ [9]} Salasnich L 2002 Laser Physics {\bf 12} 198 

\item{\ [10]} Salasnich L, Parola A, and Reatto L 2002 
Phys. Rev. A {\bf 65} 043614 

\item{\ [11]} Gross E P 1961 Nuovo Cimento {\bf 20} 454; 
Pitaevskii L P 1961 Zh. Eksp. Teor. Fiz. {\bf 40}, 
646 [English Transl. 1961 Sov. Phys. JETP {\bf 13} 451]

\item{\ [12]} Olshanii M 1998 Phys. Rev. Lett. {\bf 81} 938  

\item{\ [13]} Chiofalo M L and Tosi M P 2000 Phys. Lett. A 
{\bf 268} 406 

\item{\ [14]} Landau L D and Lifsits E M 1991 {\it Mechanics, 
Course of Theoretical Physics}, vol. 3 (Pergamon Press: Oxford); 
Arnold V I 1990 {\it Mathematical Methods of Classical Mechanics} 
(Springer: Berlin). 

\item{\ [15]} Andrews M R, Townsend C G, Miesner H J, 
Drufee D S, Kurn D M, and Ketterle W 1997 
Science {\bf 275} 637 

\item{\ [16]} Fribel S, D'Andrea C, Walz J, Weitz M, 
and H\"ansch T W 1998 Phys. Rev. A {\bf 57} R20  

\item{\ [17]} Jauregui R, Poli N, Roati G, and Modugno G 2001 
Phys. Rev. A {\bf 64} 033403  

\item{\ [18]} Kevrekidis P G, Bishop A R, and Rasmussen K O
2000 J. Low Temp. Phys. {\bf 120} 205  

\item{\ [19]} Salasnich L 2000 Phys. Lett. A {\bf 266} 187; 
Salasnich L 2000 Progr. Theor. Phys. Suppl. {\bf 139} 414 

\item{\ [20]} Cohen-Tannouji C 1998 Rev. Mod. Phys. {\bf 70} 707 

\end{description}

\newpage 

\begin{figure}
\centerline{\psfig{file=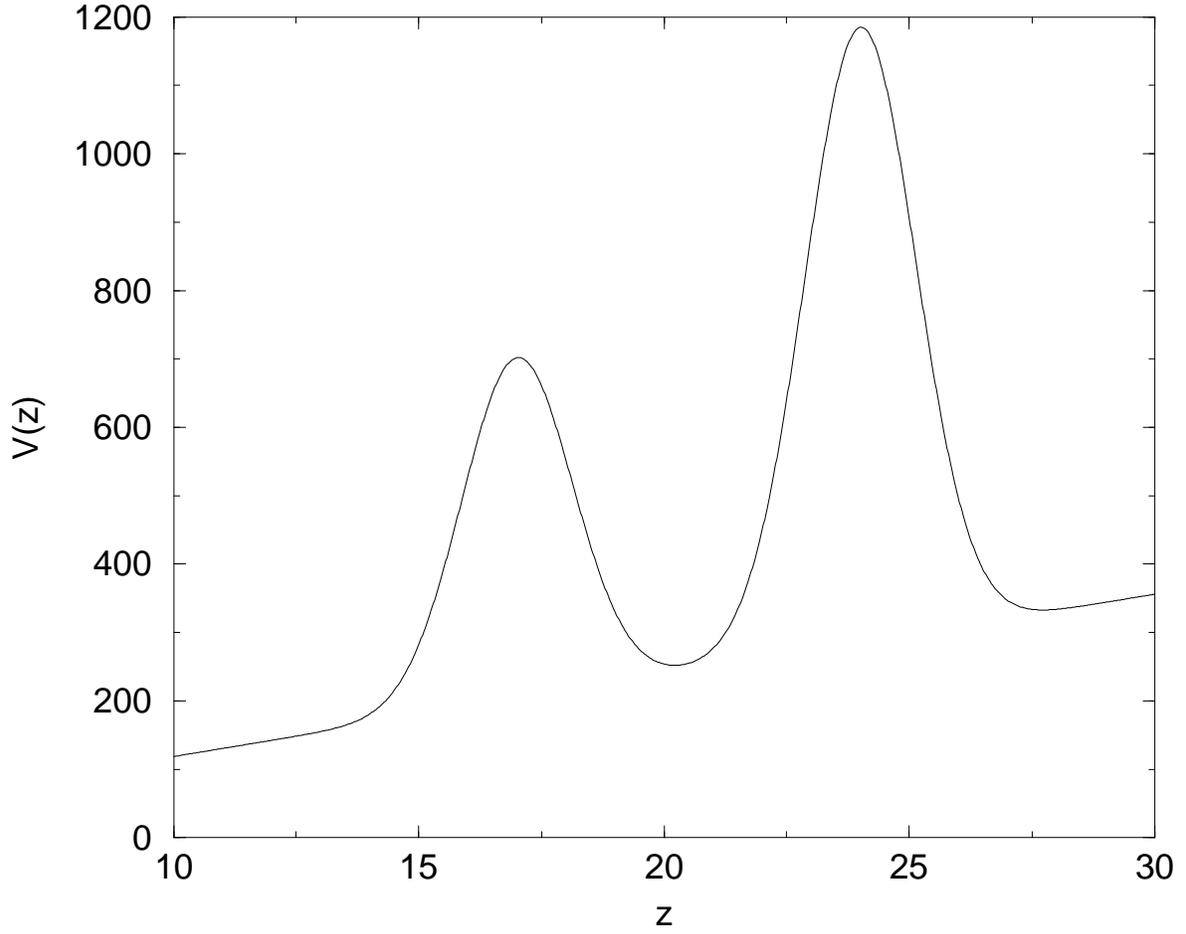,height=5.in}}
\caption{Vertical axial external potential $V(z)$ of 
Eq. (2). Scaled parameters of the potential: $V_1=300$, $V_2=900$, 
$z_1=17$, $z_2=24$, $\sigma=1.6$. 
Length $z$ in units $a_z=(\hbar/m\omega_z)$, 
where $\omega_z=\omega_{\bot}/10$ with $\omega_{\bot}=2\pi$ kHz. 
Energy in units $\hbar \omega_z$.} 
\end{figure}

\newpage 

\begin{figure}
\centerline{\psfig{file=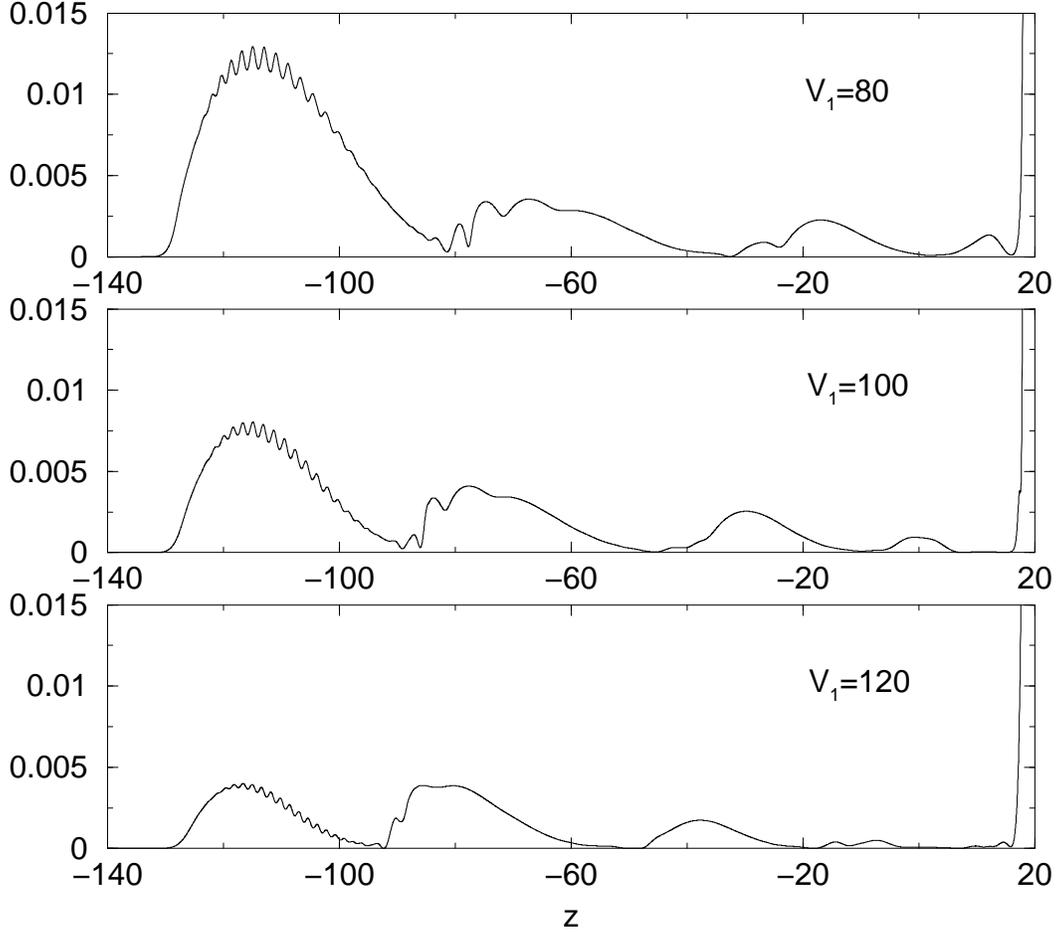,height=5.in}}
\caption{Axial density profile of 
Bose condensed $^{23}$Na atoms tunneling through 
the potential well (Eq. (5)), as obtained by solving NPSE. 
Time: $t=4.8$. Number of $^{23}$Na atoms: $N=10^4$. 
Scattering length: $a_s=30$ $\AA$. Potential well parameters: 
$z_1=17$, $z_2=24$ and $V_2=900$. Three values 
of $V_1$ correspond to three values of the 
effective chemical potential $\mu_{eff}$ and of the 
period $\tau_{osc}$ of small oscillations around the minimum of 
the potential well: $\mu_{eff}=53.91$ and $\tau_{osc}=1.21$ (top), 
$\mu_{eff}=54.23$ and $\tau_{osc}=1.17$ (middle), 
$\mu_{eff}=54.55$ and $\tau_{osc}=0.90$ (bottom).  
Length $z$ in units $a_z=(\hbar/m\omega_z)$, 
where $\omega_z=\omega_{\bot}/10$ with $\omega_{\bot}=2\pi$ kHz; 
energy in units $\hbar \omega_z$ and time in units $\omega_z^{-1}$.} 
\end{figure}
  
\newpage

\begin{figure}
\centerline{\psfig{file=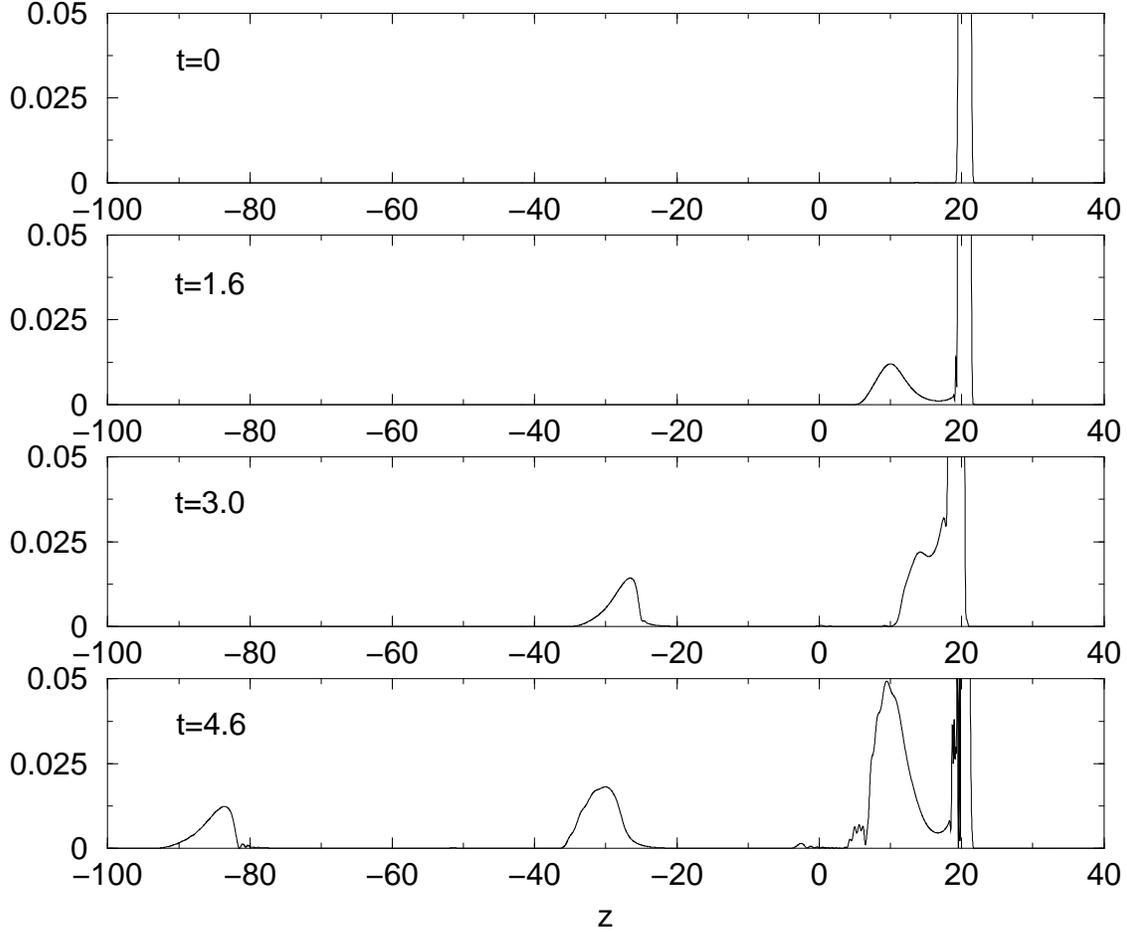,height=5.in}}
\caption{Axial density profile of Bose condensed $^{23}$Na atoms 
tunneling through the potential well 
(Eq. (2)) with oscillating barrier: 
$V_1(t)=V_a+V_b\sin{({2\pi\over \tau} t)}$, 
$V_a=300$ and $V_b=200$. Results obtained by solving NPSE. 
Period of oscillation: $\tau =1.5$. 
Effective chemical potential of the 
initial condensate: $\mu_{eff}=57.43$. 
Number of $^{23}$Na atoms: $N=10^4$. 
Scattering length: $a_s=30$ $\AA$.  
Note that $P_T=0.26$ with $V_1(t)=V_a-V_b=100$. 
Units as in Fig. 2.} 
\end{figure} 

\newpage

\begin{figure}
\centerline{\psfig{file=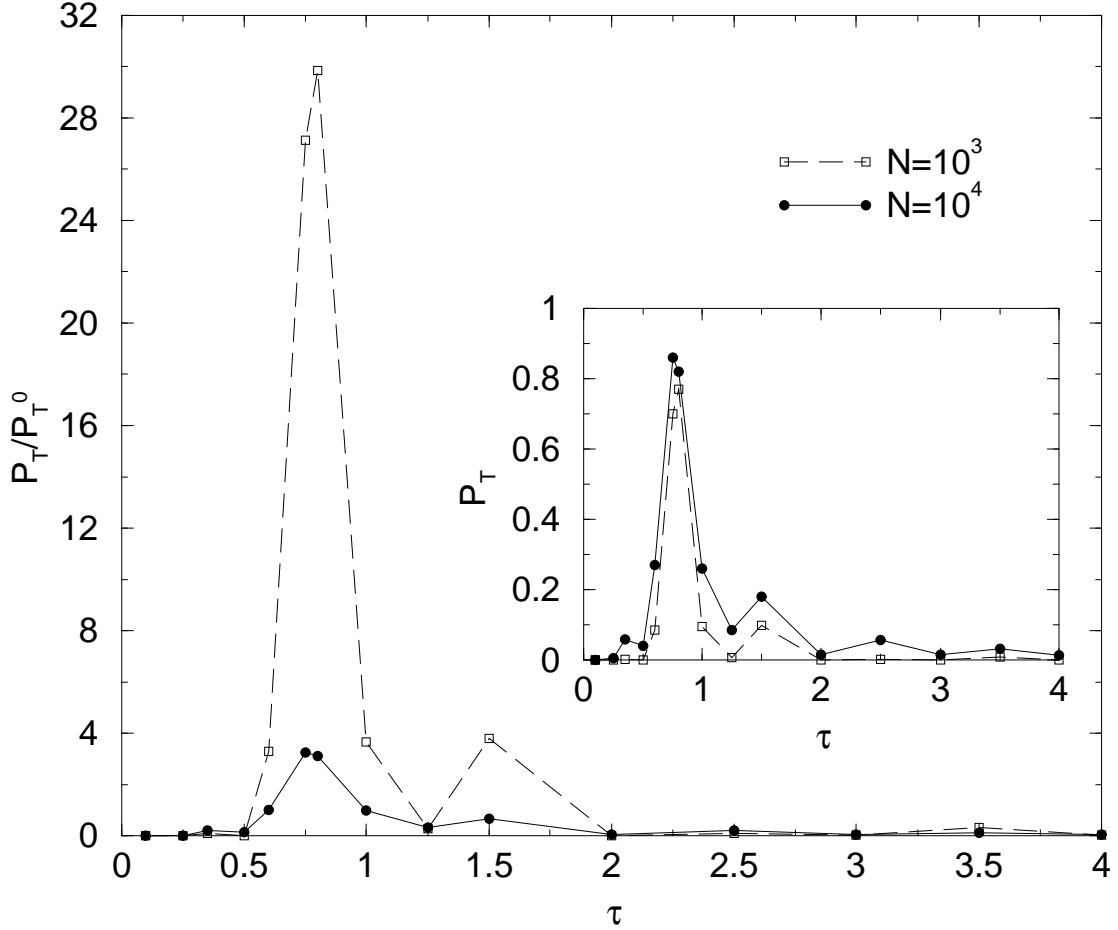,height=5.in}}
\caption{Tunneling ratio $P_T/P_T^0$ at $t=5$ as a function 
of the period $\tau$ of oscillation of the lower-lying 
Gaussian barrier (Eq. (3)), as obtained by solving NPSE. 
$V_1(t)=V_a+V_b\sin{({2\pi\over \tau} t)}$, 
$V_a=300$ and $V_b=200$ (Eq. (6)). 
The period $\tau_{osc}$ of small oscillations around the minimum of 
the potential well with $V_a=300$ and $V_b=0$ is $\tau_{osc}=0.84$. 
$P_T^0$ is the tunneling probability with $V_1=V_a-V_b=100$. 
Effective chemical potential of the 
initial condensate: $\mu_{eff}=27.15$ for $N=10^3$, 
$\mu_{eff}=57.43$ for $N=10^4$. 
$N$ is the number of $^{23}$Na atoms. 
Scattering length: $a_s=30$ $\AA$. Inset: tunneling probability 
$P_T$ at t=5 as a function of $\tau$. Units as in Fig. 2.} 
\end{figure} 

\newpage

\begin{figure}
\centerline{\psfig{file=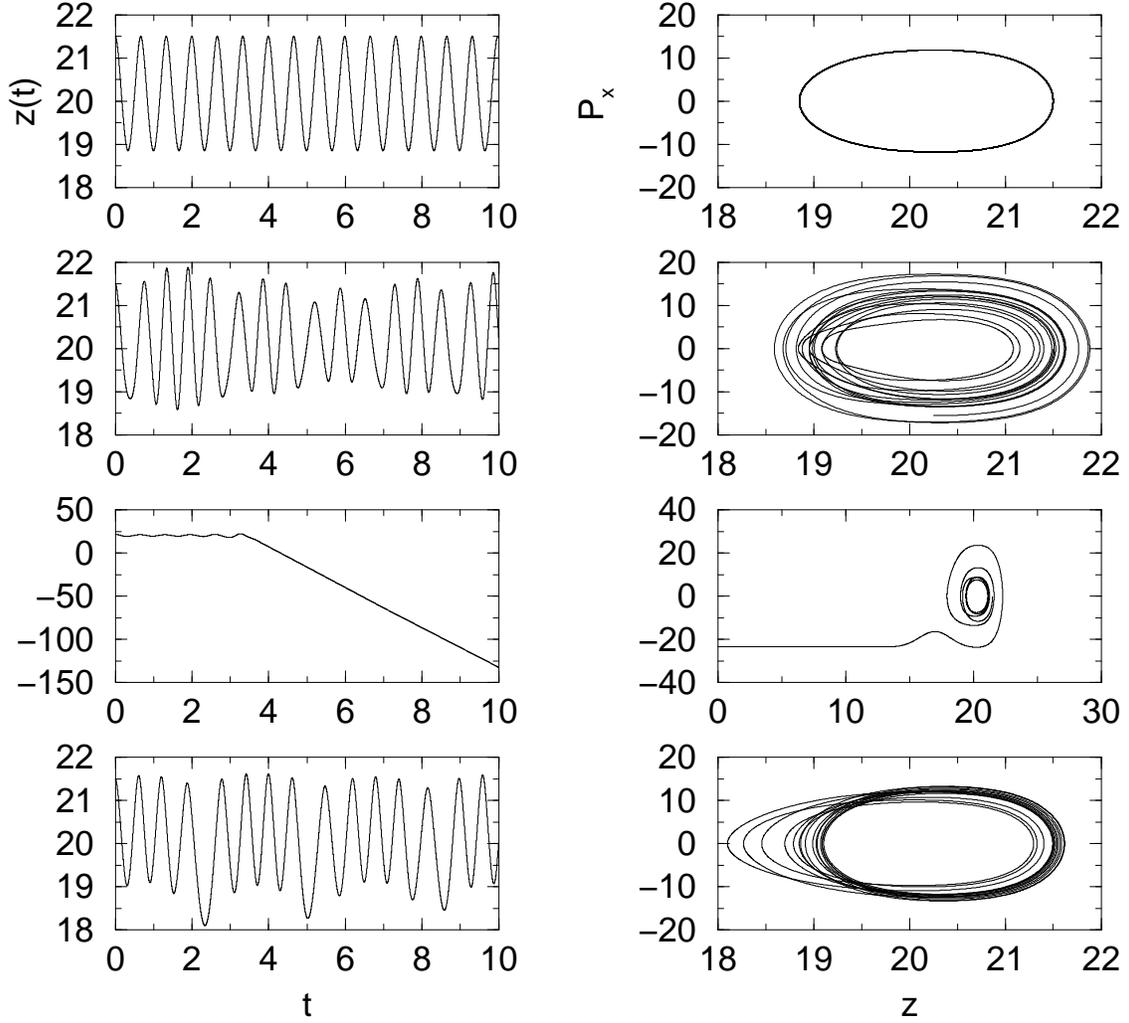,height=5.5in}}
\caption{Axial coordinate $z(t)$ (left) and phase-space portrait $(z,p_z)$ 
with $p_z=\dot z$ (right) of a classical particle under 
the action of the potential $V(z)$ given by Eq. (2) with 
its parameters as in Fig. 4. From top to bottom: 
(a) $V_b=0$ and $\tau=\;$anything; (b) $V_b=200$, 
$\tau=0.5$; (c) $V_b=200$ and $\tau=0.75$; 
(d) $V_b=200$ and $\tau=3$. Units as in Fig. 2.} 
\end{figure} 
      
\newpage

\begin{figure}
\centerline{\psfig{file=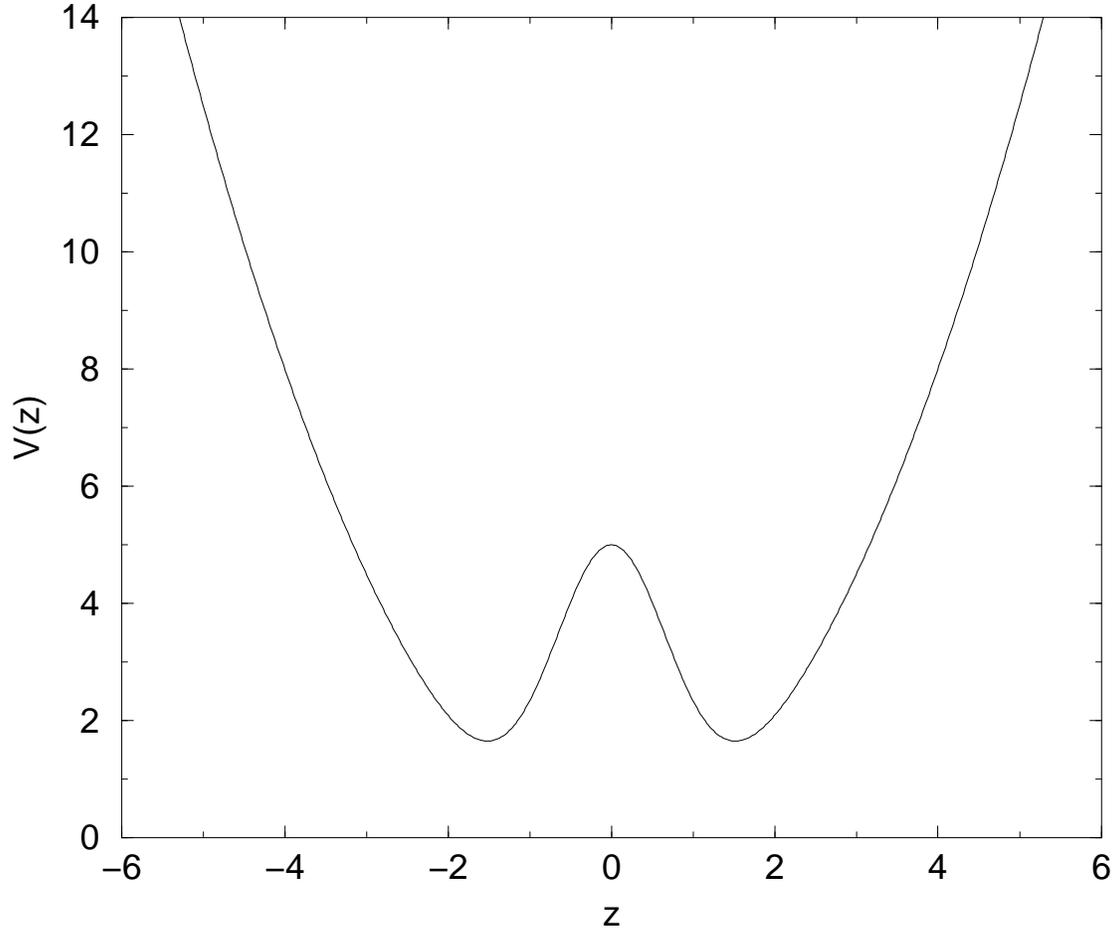,height=5.in}}
\caption{Horizontal axial double-well potential 
$V(z)$ given by Eq. (4). Scaled parameters of the potential: 
$V_1=5$, $V_2=0$, $\sigma=1$.
Length $z$ in units $a_z=(\hbar/m\omega_z)$ 
and energy in units $\hbar \omega_z$.} 
\end{figure} 

\newpage

\begin{figure}
\centerline{\psfig{file=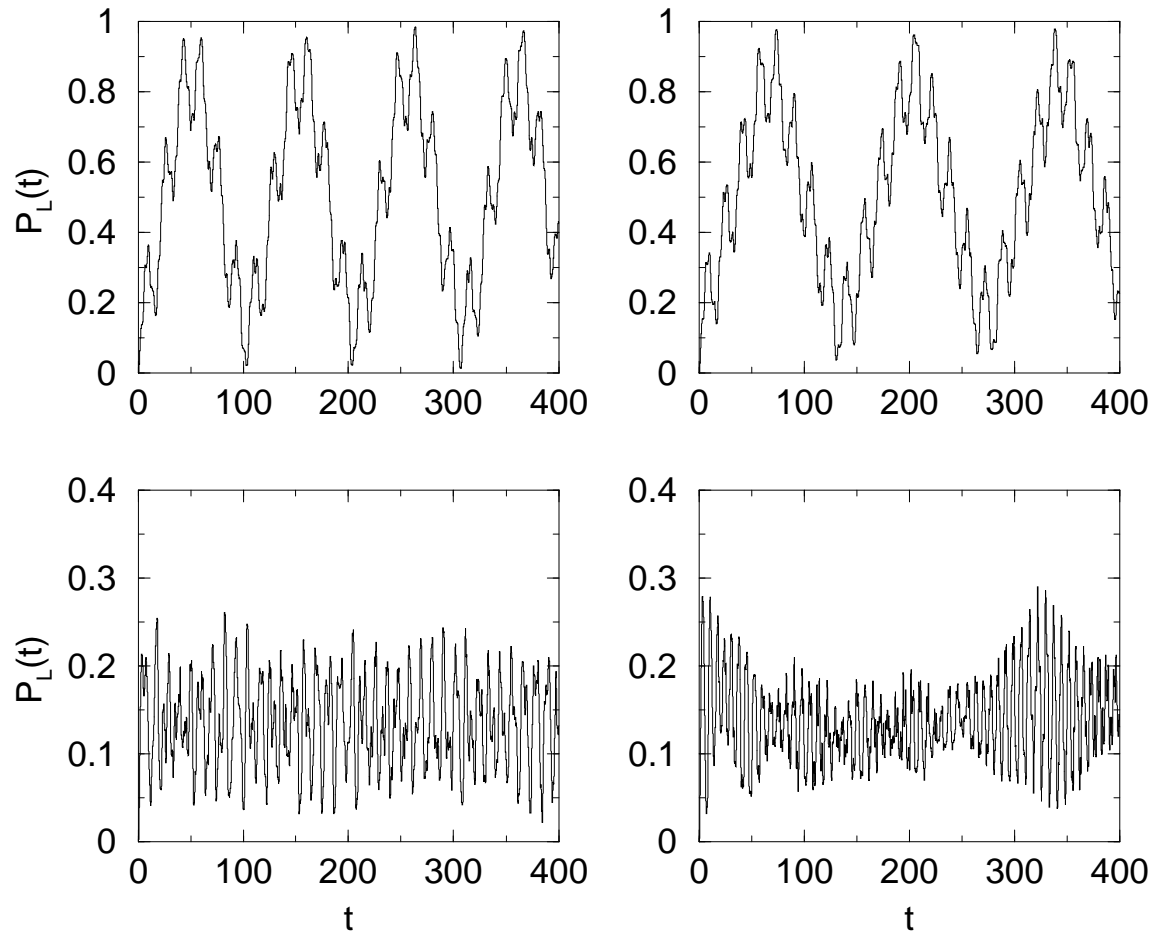,height=5.in}}
\caption{Fraction $P_L(t)$ of Bose condensed $^{23}$Na atoms 
in the left well obtained by solving NPSE 
where the external potential is 
given by Eq. (3) with $V_1=6.5$ and $V_2=0$. 
From top to bottom and from left to right: 
(a) $Na_s/a_z=0$; (b) $Na_s/a_z=0.014$; 
(c) $Na_s/a_z= 0.07$; (d) $Na_s/a_z=0.14$. 
Length $z$ in units $a_z=(\hbar/m\omega_z)$, 
where $\omega_z = \omega_{\bot}/10$ with 
$\omega_{\bot}=2\pi$ kHz. Energy in units 
$\hbar \omega_z$ and time in units $\omega_z^{-1}$.} 
\end{figure} 

\newpage

\begin{figure}
\centerline{\psfig{file=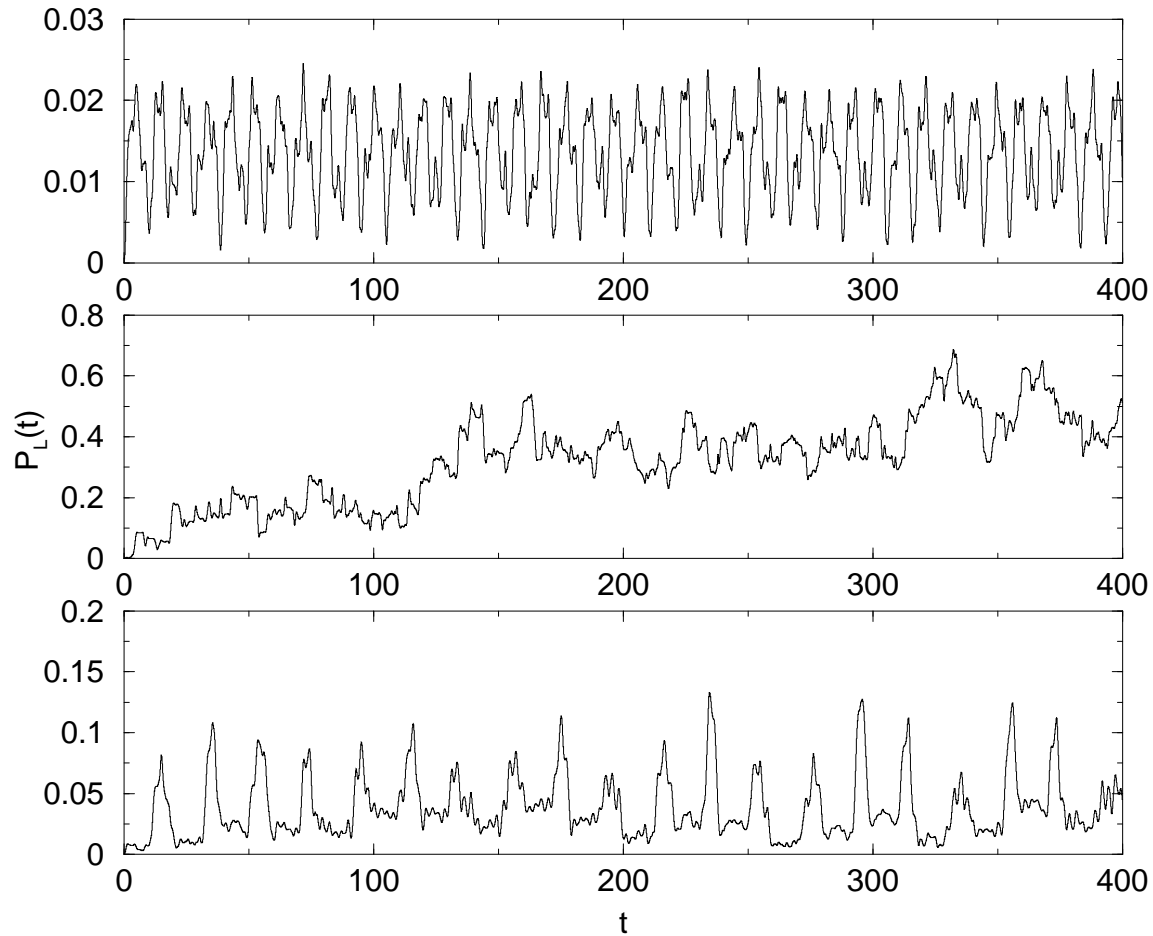,height=5.in}} 
\caption{Fraction $P_L(t)$ of Bose condensed $^{23}$Na atoms 
in the left well obtained by solving NPSE 
where the external potential is given by Eq. (3) 
with $V_1=6.5$. From top to bottom: 
(a) $V_2=0$ and $\tau=\;$anything; (b) $V_2=2$ and $\tau = \tau_{ST}/2$; 
(c) $V_2=2$ and $\tau = 2 \tau_{ST}$; where $\tau_{ST}$ is the period 
of unperturbed MQST oscillation. Units as in Fig. 7.} 
\end{figure} 

\newpage

\begin{figure}
\centerline{\psfig{file=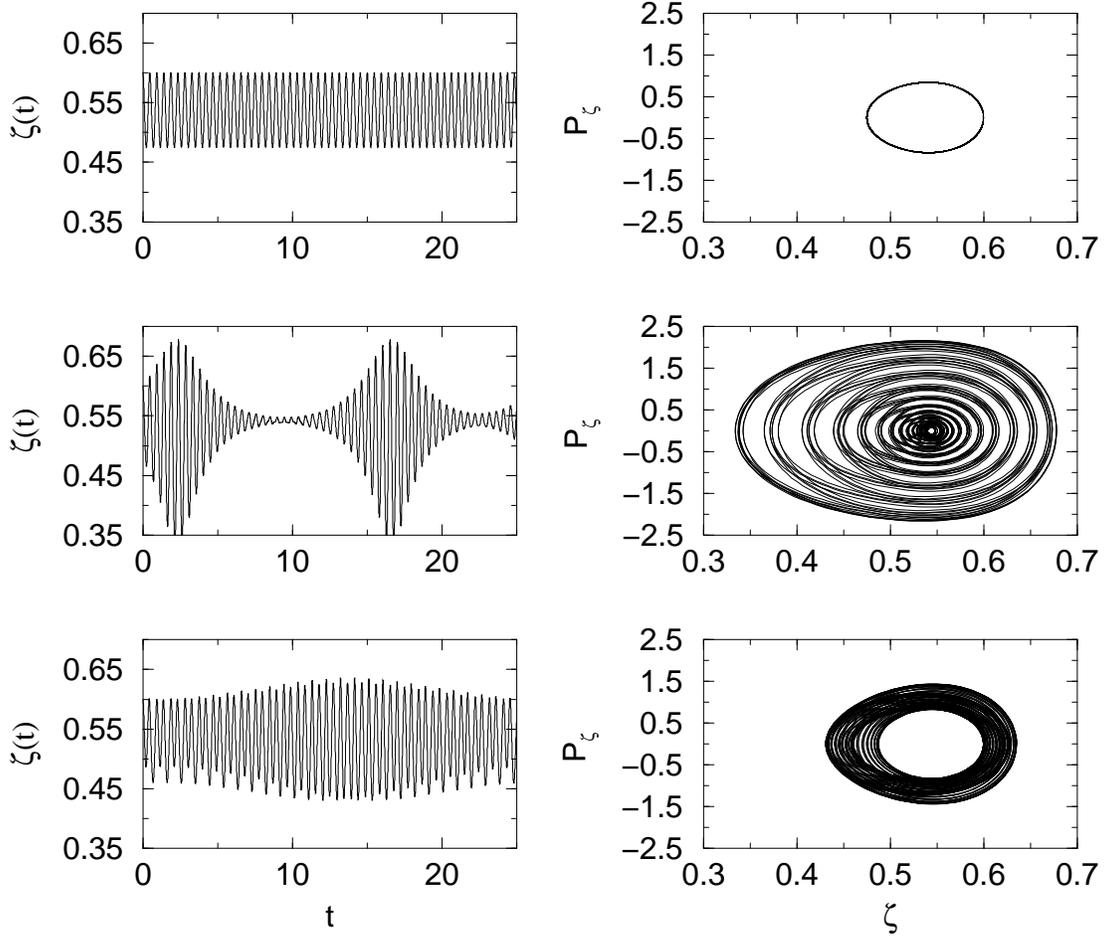,height=5.in}}
\caption{Fractional population 
imbalance $\zeta(t)$ of the condensate 
in the two wells (left) and its phase-space portrait 
$(\zeta,p_{\zeta})$ with $p_{\zeta}={\dot \zeta}$ (right), 
as obtained by solving Eq. (6). 
Initial conditions: 
$\zeta(0)=0.6$, $\phi(0)=0$. 
$\Lambda=\Lambda_0(1+\epsilon \sin{(2\pi t/\tau)})$ 
where $\Lambda_0=25$. From top to bottom: 
(a) $\epsilon = 0$ and $\tau=\;$anything; (b) $\epsilon = 0.1$ and 
$\tau= \tau_{ST}/2$; (c) $\epsilon =0.1$ and $\tau = 2 \tau_{ST}$;  
where $\tau_{ST}$ is the period of unperturbed MQST oscillations. } 
\end{figure} 

\newpage

\begin{figure}
\centerline{\psfig{file=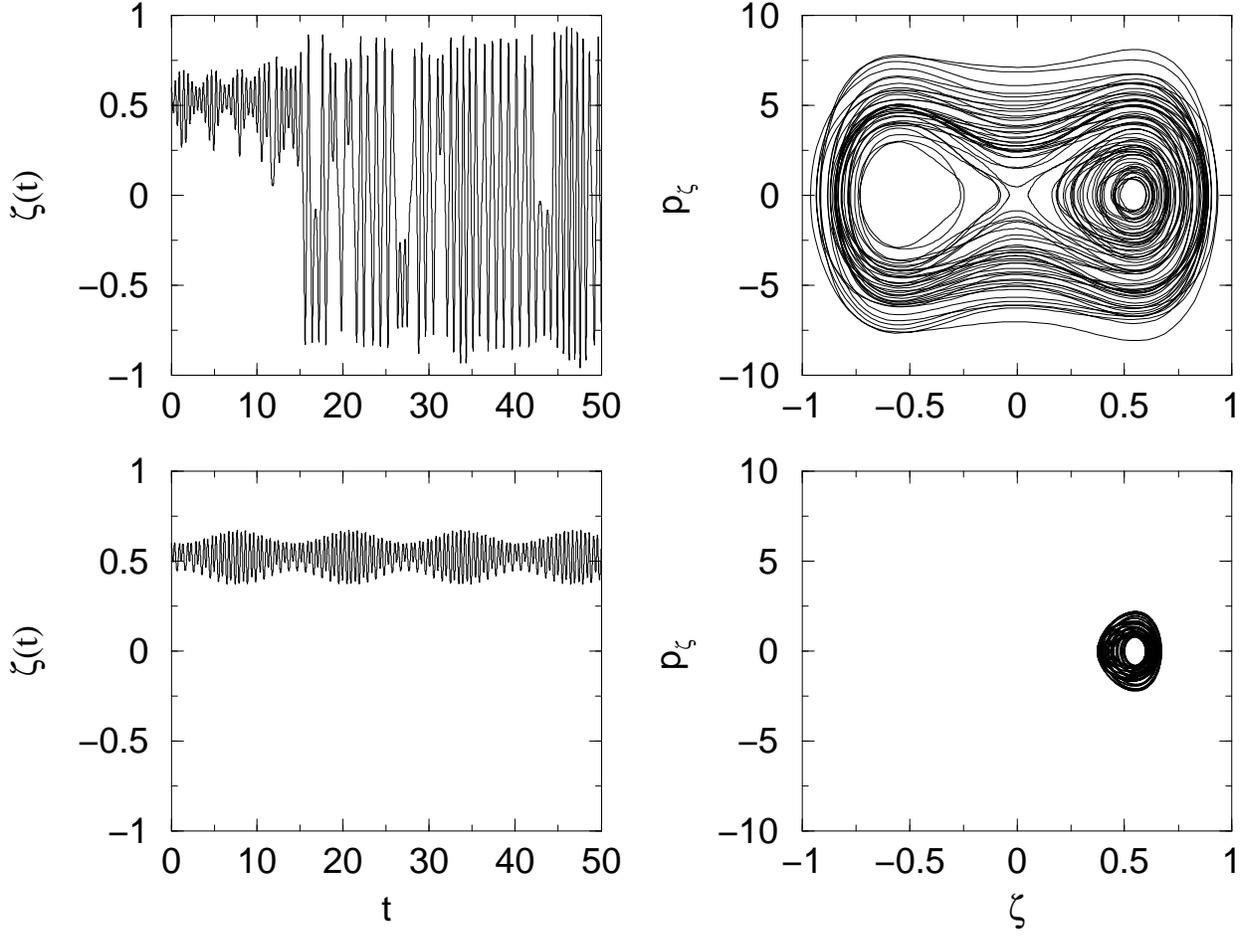,height=5.in}}
\caption{Fractional population imbalance $\zeta(t)$ of the condensate 
in the two wells (left) and its phase-space portrait 
$(\zeta,p_{\zeta})$ with $p_{\zeta}={\dot \zeta}$ (right), 
as obtained by solving Eq. (6). 
Initial conditions: $\zeta(0)=0.6$, $\phi(0)=0$. 
$\Lambda=\Lambda_0(1+\epsilon \sin{(2\pi t/\tau)})$ 
where $\Lambda_0=25$ and $\epsilon = 0.2$. From top to bottom: 
(a) $\tau = \tau_{ST}/2$; (b) $\tau=2 \tau_{ST}$; where 
$\tau_{ST}$ the period of unperturbed MQST oscillations. } 
\end{figure} 
        
\end{document}